\documentstyle[12pt]{article}
\begin{document}
\hyphenation{e-qui-va-lent}
\hyphenation{nee-ded}
\hyphenation{eu-ler}
\hyphenation{va-ria-tio-nal}
\hyphenation{e-qua-tion}
\hyphenation{fad-deev-po-pov}
\def \KG{\hbox {{$ \Box $} \hskip-12.5pt/ }}
\def \Fs{F^{\mu \nu}}
\def \Fg{F_{\mu \nu}}
\def \Rs{R^{\mu \nu}}
\def \Rg{R_{\mu \nu}}
\begin{titlepage}
\title{\LARGE Gauge fixing in Higher Derivative Field Theories
\thanks{Submitted to Nuclear Physics B.} \\
\author{A.Bartoli \\
Dipartimento di Fisica, Universit\`a di Bologna, \\
I.N.F.N. Sezione di Bologna, Via Irnerio 46, Bologna Italy.
\and J. Julve \thanks{Partially supported by DGICYT and the CNR-CSIC
cooperation agreement.} \\
Instituto de Matem\'aticas y F\'{\i}sica Fundamental, \\
CSIC, Serrano 123, Madrid, Spain. }}
\maketitle
\begin{abstract}
Higher Derivative (HD) Field Theories can be transformed into
second order equivalent theories with a direct particle
interpretation. In a simple model involving abelian gauge
symmetries we examine the fate of the possible gauge fixings
throughout this process. This example is a useful test bed for HD
theories of gravity and provides a nice intuitive interpretation
of the "third ghost" occurring there and in HD gauge theories
when a HD gauge fixing is adopted.
\end{abstract}
\vfill
Preprint IMAFF 93/10 \\
\end{titlepage}

\section*{Introduction}
\parindent 1 cm
Higher Derivative (HD) field theories arise as effective theories
in several contexts. Perhaps, the best known example is
gravitation, where higher order terms in the curvatures arise
from an underlying fundamental string dynamics or from quantizing
matter fields in a curved space-time background.
The study of the actual dynamical degrees of freedom
(d.o.f) of such theories has been faced most fruitfully by
bringing them to a lower derivative (LD) equivalent version of
second order by means of a Legendre transformation \cite{1}.
More recently, progress has been made towards a complete
diagonalization of these d.o.f in models with quadratic terms in
the scalar and Ricci curvatures \cite{2}.\par
However in all these studies only invariant theories under
general coordinate transformations have been considered which,
as such, are unsuited for quantization. In fact, the Green's
functions for the equations of motion are undefined
because of the Diff-invariance. The way out of this difficulty in
gauge theories is to break the local symmetry by introducing a
Gauge Fixing (GF) term, which can be now of the LD or HD type
according to computational convenience.
Beside the usual gauge ghosts, HD gauge fixings introduce a
more subtle "third ghost" \cite{3}, which calls for
further compensa\-ting Faddeev-Popov ghosts. Once the GF term has
been added to the HD theory the question arises of how does it
translate through the Legendre transform down to the LD
equivalent theory where the particle contents is apparent. This
may also provide an intuitive picture for the third ghost.\par
In this paper we explore these questions by using a simple HD
abelian gauge theory as a test bed. In Section 1 we introduce the
model and point out the nature of the states it contains by
studying the (tree approximation) propagator. We then perform the Legendre
transform in Section 2 to end up with the equivalent Helmholtz LD
theory. In Section 3 we test the reliability of the formal covariant
treatment by working out explicitely the actual d.o.f of the radiating
field. In section 4 we outline the fate of the gauge
symmetry(ies) along the process, which is more intriguing when
only a LD GF is adopted. Finally we draw some conclusions and
consequences of this study. Some details complementary to ref.[4]
are given in an Appendix together with the notations used.
 Through this paper 4D space-time with
Lorentzian signature (+ - - - ) is considered.\par

\section{The Model}
We consider the simplest fourth-derivative abelian gauge theory
quadratic in the gauge fields obtained by extrapolating the QED
Lagrangian in a natural way
\begin{equation}
{\cal L}_{HD} = - \frac{1}{4} \Fg \Fs - \frac
{1}{4m^2} \Fg \Box \Fs - \frac{\zeta^2}{2} ( \partial_{\mu}
A^{\mu} )^2 - \frac{\zeta^2}{2M^2} ( \partial_{\mu} A^{\mu} )
\Box ( \partial_{\nu} A^{\nu} ) - j_{\mu} A^{\mu}
\end{equation}
where $ \Fg = \partial_{\mu} A_{\nu} - \partial_{\nu} A_{\mu} $.
Here $ m^2 $ is a dimensional parameter, $\zeta$ and $M^2$ are
dimensionless and dimensional gauge parameters respectively.\par For
$\zeta = 0 $ the theory is invariant under the U(1) gauge
transformations
\begin{equation}
\delta A_{\mu} = \partial_{\mu} \lambda (x)
\end{equation}
provided that the source $j^{\mu}$ is conserved, namely if
\begin{equation}
\partial_{\mu} j^{\mu} = 0
\end{equation}
The first term in (1) is reminiscent of $\sqrt {-g}R $ in the
gravity case, whereas the second term is reminescent of $\sqrt
{-g}R^2 $ and $\sqrt {-g} \Rg \Rs $ (usually a term of the form $\sqrt
{-g} R^{\alpha}_{\mu \beta \nu} R_{\alpha}^{\mu \beta \nu} $ is
not considered because of the Gauss-Bonnet identity).\par
Though consisting of a decoupled sector in the Abelian case, it may be
instructive to consider the Faddeev-Popov (FP) Lagrangian ${\cal L}_{FP}$
for the gauge fixing in (1). Together with the gauge-breaking terms it can
be expressed as a coboundary in the BRS cohomology
\begin{equation}
{\cal L}_g = \bar \delta [\bar c (1+ \frac {\Box}{M^2} ) \partial_{\mu}
A^{\mu} + \frac {1}{2 \zeta^2} \bar c ( 1+ \frac {\Box}{M^2} ) B']
\end{equation}
where the BRS transformation $\bar \delta$ is defined by
\begin{eqnarray*}
\bar \delta A_{\mu} & = & \partial_{\mu} c \\
\bar \delta c & = & 0 \\
\bar \delta \bar c & = & B' \\
\bar \delta B' & =& 0
\end{eqnarray*}
Ghost numbers and mass dimensions are the usual ones: $$ gn(c)=-gn(\bar c)=1,
 gn(A)=gn(B)=0;\quad [c]=0, [A]=1, [\bar c]=[B]=2.$$
Upon the redefinition of the auxiliary field
$$B'= B - \zeta^2 \partial_{\mu} A \quad , $$
Eq.(4) gets the usual diagonalized form
\begin{equation}
{\cal L}_g = - \frac{\zeta^2}{2} (\partial_{\mu}
A^{\mu}) (1+ \frac {\Box}{M^2} ) (\partial_{\nu}
A^{\nu}) -  \bar c ( 1+ \frac {\Box}{M^2} ) \Box c + \frac{1}{2 \zeta^2}
B ( 1+ \frac {\Box}{M^2} ) B
\end{equation}
where the first term yields the gauge-breaking part in (1),
and the last two will be referred to as ${\cal L}_{FP}$ and ${\cal
L}_B$ respectively in the
following. We now go on with the sector (1) of the theory and
come on ${\cal L}_{FP}$ and ${\cal L}_B$ later. \par
Dropping total derivatives, (1) can be written in the more
convenient form
\begin{equation}
{\cal L}_{HD}  = \frac {1}{2m^2} A^{\mu} \Box ( \Box + m^2 )
\theta_{\mu \nu} A^{\nu} + \frac {\zeta^2}{2M^2}
A^{\mu} \Box ( \Box + M^2 ) \omega_{\mu \nu} A^{\nu} - j_{\mu} A^{\mu}
\end{equation}
where
\begin{eqnarray}
\omega_{\mu \nu} &= & \frac{\partial_{\mu} \partial_{\nu}}{\Box}
\nonumber \\
\theta_{\mu \nu} &= & \eta_{\mu \nu} - \frac{\partial_{\mu}
\partial_{\nu}}{\Box}
\end{eqnarray}
are the longitudinal and transverse projectors respectively.
They are a complete set of orthogonal projectors in the gauge
field space. Their properties
\begin{eqnarray}
\theta_{\mu \nu} +  \omega_{\mu \nu} &= & \eta_{\mu \nu}
\nonumber \\
\theta_{\mu \rho} \theta^{\rho}_{ \nu} &= & \theta_{\mu \nu}
\nonumber \\
\omega_{\mu \rho} \omega^{\rho}_{ \nu} &= & \omega_{\mu \nu}
\nonumber \\
\theta_{\mu \rho} \omega^{\rho}_{ \nu} &= & 0
\end{eqnarray}
are fully exploited in what follows.\par
The form given in (6) is specially suited to work out the
gauge field propagator which formally reads
\begin{equation}
\Delta_{\mu \nu} = \theta_{\mu \nu} \frac{m^2}{\Box ( \Box +
m^2)} + \omega_{\mu \nu} \frac{1}{\zeta^2}  \frac{M^2}{\Box (
\Box + M^2)}
\end{equation}
and lends itself to a direct reading of the particle contents. In
fact (9) may be rewritten as
\begin{equation}
\Delta_{\mu \nu} = \theta_{\mu \nu} \left( \frac{1}{\Box} -
\frac{1}{\Box + m^2} \right) + \omega_{\mu \nu} \frac{1}{\zeta^2}
\left( \frac{1}{\Box} - \frac{1}{\Box + M^2} \right).
\end{equation}
Thus the theory propagates a transverse massless vector field (a
photon accounting for 2 d.o.f) and a massive negative-norm
transverse vector (a "poltergeist" with 3 d.o.f). Beside this,
one has a massless longitudinal d.o.f (the "gauge ghost") and a
massive longitudinal "poltergeist" d.o.f (the "third ghost").\par
Unlike the second order theories, we see from (5) that also the FP
ghosts $\bar c$, $c$ actually propagate {\it two} d.o.f:
$$ \Delta_c = \frac {1}{\Box} - \frac {1}{\Box + M^2} \quad , $$
and the auxiliary field B now describes a {\it propagating} massive
d.o.f
$$ \Delta_B = \frac{\zeta^2 M^2}{\Box + M^2} \quad . $$  \par
The aim of the Legendre transform is to provide a 2nd-derivative
equivalent theory with explicit independent fields for the
positive-norm and the poltergeist d.o.f's. Before doing this we
simplify the notation by omitting the indices and write (6) as
\begin{equation}
{\cal L}_{HD}  =  \frac {1}{2} A \Box (\theta + \zeta^2 \omega
)A + \frac {1}{2} A \Box \Box ( \frac{\theta}{m^2} +
\frac{\omega}{M^2})(\theta + \zeta^2 \omega )A - jA
\end{equation}
By defining
\begin{equation}
\hat A = (\theta + \zeta \omega )A
\end{equation}
and making further use of (8), Eq.(11) can be brought to the
final condensed form
\begin{equation}
{\cal L}_{HD}[\hat A , \Box \hat A ]  =  \frac {1}{2} \hat A
\Box \hat A + \frac {1}{2} \hat A \Box \Box ( \frac{\theta}{m^2}
+ \frac{\omega}{M^2})\hat A - j (\theta + \frac{\omega}{\zeta} ) \hat A.
\end{equation}

\section {Legendre Transform and Helmholtz Lagrangian }
In a general HD theory ${\cal L} [\varphi, \partial \varphi]$
the problem arises of finding a function $f[\partial
\varphi]$ of field derivatives of various orders which is
suitable to define a canonical conjugate variable
\begin{equation}
\pi = \frac{\partial{\cal L}}{\partial f [\partial \varphi]} .
\end{equation}
The condition that this equation be invertible (hyper-regu\-lar
systems), namely that $ f [\partial \varphi]$ can be worked out
as a function of $\varphi$ and $\pi$, usually allows just one
choice for $ f [\partial \varphi]$. In our case the unambiguous
choice is the object $  \Box \hat A $ and (13) has been
prepared accordingly. One finds
\begin{eqnarray}
\hat\pi &= &\frac{\partial{\cal L}}{\partial \Box \hat A}
\nonumber \\
& = &  \frac{1}{2} \hat A + ( \frac{\theta}{m^2} +
\frac{\omega}{M^2}) \Box \hat A
\end{eqnarray}
from which
\begin{equation}
\Box \hat A = (m^2 \theta + M^2 \omega )(\hat \pi - \frac{1}{2}
\hat A ) \equiv F[\hat A, \hat \pi]
\end{equation}
The Hamiltonian function is then
\begin{eqnarray}
{\cal H} [\hat A, \hat \pi] &= & \hat A F[\hat A, \hat \pi] -
{\cal L}_{HD}[\hat A , F[\hat A, \hat \pi] ] \nonumber \\
&= & \frac{1}{2} (\frac{1}{2} \hat A - \hat \pi ) (m^2 \theta +
M^2 \omega )(\frac{1}{2} \hat A - \hat \pi ) + j (\theta +
\frac{\omega}{\zeta} ) \hat A
\end{eqnarray}
The canonical equations of motion for $\hat A$ and $\hat \pi $
are the system of (2nd order) equations
\begin{eqnarray}
\Box \hat A &= & \frac{\partial {\cal H}}{\partial \hat \pi}
\nonumber \\
\Box \hat \pi &= & \frac{\partial {\cal H}}{\partial \hat A }
\end{eqnarray}
which is equivalent to the Euler's equation from (13).
However both Eqs.(18) can also be derived by a variational
principle from the so called Helmholtz Lagrangian
\begin{equation}
{\cal L}_H [\hat A, \hat \pi] = \hat \pi \Box \hat A - {\cal H}
[\hat A, \hat \pi]
\end{equation}
It depends on $\hat A, \hat \pi $ and derivatives of $\hat A $.
The first (derivative) term looks like a kinetic one while the
terms in ${\cal H} [\hat A, \hat \pi] $ are of the
mass-term type, the problem being that $\hat A$ and $\hat \pi$
occur mixed. The diagonalization can be trivially performed by
defining new tilde fields such that
\begin{eqnarray}
\hat A &= & \tilde A + \tilde \pi \nonumber \\
\hat \pi &= & \frac{1}{2} ( \tilde A - \tilde \pi )
\end{eqnarray}
Eq.(19) becomes
\begin{eqnarray}
{\cal L}_H [\tilde A, \tilde \pi] & = & \frac{1}{2} \tilde A \Box
\tilde A - \frac{1}{2} \tilde \pi [( \Box + m^2) \theta + ( \Box
+ M^2) \omega ] \tilde \pi \nonumber \\
& & \mbox{} - j (\theta + \frac{\omega}{\zeta} ) (\tilde A +
\tilde \pi )
\end{eqnarray}
In terms of fields that couple directly to the source, namely
\begin{eqnarray}
{\cal A} &= & (\theta + \frac{\omega}{\zeta} ) \tilde A \nonumber
\\
{\Pi} &= & (\theta + \frac{\omega}{\zeta} ) \tilde \pi
\end{eqnarray}
Eq.(21) finally gives the desired LD theory
\begin{eqnarray}
{\cal L}_{LD} [{\cal A} , {\Pi} ] & = & \frac{1}{2} {\cal A}
\Box \theta {\cal A} - \frac{1}{2} {\Pi} ( \Box + m^2)
\theta {\Pi} + \frac{\zeta^2}{2}  {\cal A} \Box \omega {\cal A} -
\frac{\zeta^2}{2} {\Pi} ( \Box + M^2) \omega {\Pi}
\nonumber \\
& & \mbox{} - j ({\cal A} + {\Pi} )
\end{eqnarray}
The physical meaning is now apparent. Whenever the source
emitted a ``particle" $A_{\mu}$ with propagator (10) (or the
effective quartic version (9)), on the same line it actually emits a massless
transverse particle ${\cal A}_{\mu} $ with propagator
\begin{equation}
\theta_{\mu \nu} \frac{1}{\Box} \quad ,
\end{equation}
a massless longitudinal (gauge ghost) state of ${\cal A}_{\mu} $ with
propagator
\begin{equation}
\omega_{\mu \nu } \frac{1}{\zeta^2} \frac{1}{\Box}
\end{equation}
{\it and}  a massive transverse poltergeist ${\Pi}_{\mu} $ with
propagator
\begin{equation}
- \theta_{\mu \nu} \frac{1}{\Box + m^2}
\end{equation}
together with a massive longitudinal ghost state (or "third
ghost") of ${\Pi}_{\mu} $ with propagator
\begin{equation}
- \omega_{\mu \nu } \frac{1}{\zeta^2} \frac{1}{\Box + M^2} \qquad .
\end{equation}
All of this amounts to the ``joint" propagator in (10).\par
The better asymptotic behaviour of the propagator (9) shows
also that the poltergeists can be viewed as regulator fields for
an otherwise LD gauge theory.\par
One should finally notice that the LD Lagrangian (23) contains
the non-local term
\begin{equation}
\frac{1}{2} (m^2 - \zeta^2 M^2 ) {\Pi}_{\mu}
\frac{\partial_{\mu} \partial_{\nu} }{\Box} {\Pi}_{\nu}
\end{equation}
which can be made to vanish by suitably choosing the gauge
parameters .\par
\vskip 1 cm
Let us now come to ${\cal L}_{FP}$ and ${\cal L}_B$. While the latter is
already of second order, ${\cal L}_{FP}$ is higher-derivative and would
in principle deserve the same treatment above. However, $\bar c$ and $c$
being independent fields, a Legendre transform cannot be carried out. In
any case the eventual diagonalization of the d.o.f described by
${\cal L}_{FP}$ is irrelevant in Abelian theories where the FP ghosts are
decoupled from the physical sector. Even in the non-Abelian case the
massive FP d.o.f does not couple to the physical sector as long
as gauge-breaking terms of the type displayed in (1) are considered. Infact
${\cal L}_{FP}$ is then
$$ - \bar c  (1+ \frac{\Box}{M^2} ) \partial_{\mu} D^{\mu} c $$
and the (field-independent) operator $ (1+ \frac{\Box}{M^2} ) $ can be
absorbed by a redefinition of the antighost $\bar c$ , factorizing a
constant functional determinant in Path Integral quantization.\par
However we may consider more general gauges of the form
\begin{equation}
(\partial_{\mu} A^{\mu}) (1+ \frac {\Box}{M^2} + f(A)) (\partial_{\nu}
A^{\nu})
\end{equation}
where $f(A)$ is a function of the quantum gauge field and/or generally
of background fields, so that also the massive FP ghost d.o.f couples to
the gauge field. Then the (propagating) auxiliary field B gets coupled
to both the gauge and the FP fields as well.

\section{Physical degrees of freedom}
We perform here a canonical analysis of the phase space
along the lines of ref.[4].\par
We consider now the non FP sector of the model and assume the conservation
of the matter source. Then (13) describes a higher-derivative theory for the
four functional d.o.f in $\hat A$ (notice that they are the same d.o.f
contained in $A$ as long as $\zeta \not= 0$).\par
Already in the higher derivative version the theory can be seen to contain
less physically meaningful configuration
d.o.f than the eight d.o.f one could expect
from the doubling (caused by the fourth differential
order of the theory) of the four quoted above. \par
Consider first the equation of motion stemming from (1), namely
\begin{equation}
\Box ( 1 + \frac{\Box}{ M^2}) A^{\mu} - [(1 - \zeta^2) + \Box
( \frac{1}{ m^2} - \frac{\zeta^2}{M^2}) ] \partial^{\mu} ( \partial_{\nu}
A^{\nu} ) = j^{\mu}
\end{equation}
Taking the divergence of both sides one gets
\begin{equation}
\Box ( 1 + \frac{\Box}{ M^2}) \partial_{\nu} A^{\nu} = 0
\end{equation}
which shows that $  \partial_{\nu} A^{\nu} $ actually describes
{\it two} decoupled free scalars (one massless and one with mass M),
already identified in (10). They amount to two configuration
variables or equivalently to four phase-space variables.\par
On the other hand, a further (non-covariant) d.o.f can be absorbed by
a redefinition of the matter fields. This can be seen by writing (1) in an
equivalent lower-order form in phase-space variables
which is the analogous of the first order (43) for the (second order) QED.
To simplify matters we get rid of the scalar d.o.f above by considering
only the Lorentz-transverse part of $A$. This is accomplished by omitting
the gauge fixing terms and remembering that $ \partial_{\nu} A^{\nu} = 0 $
when necessary. Thus we consider the Lagrangian
\begin{eqnarray}
{\cal L}^{(4)} & = & - \frac{1}{4} \Fg \KG \Fs - j_{\mu} A^{\mu}
\nonumber \\
& = & \frac{1}{2} ( - \partial_0 \vec A - \vec \nabla A^0 ) \KG
( - \partial_0 \vec A - \vec \nabla A^0 ) - \frac{1}{2} (\vec \nabla \times
\vec A ) \KG (\vec \nabla \times \vec A ) \nonumber \\
& & \mbox{} - j_{\mu} A^{\mu}
\end{eqnarray}
where $ \KG$ stands for the Klein-Gordon operator
$  ( 1 + \frac{\Box}{ m^2}) $  and the remaining
notations are given in the Appendix.\par
Then the lower-order Lagrangian is
\begin{equation}
{\cal L}^{(3)} = \vec E \KG ( - \partial_0 \vec A - \vec \nabla A^0 )
- \frac{1}{2} [ \vec E \KG \vec E + (\vec \nabla \times
\vec A ) \KG (\vec \nabla \times \vec A ) ] - A^0 \rho + \vec A \vec j
\end{equation}
Notice that solving the equation of motion for $\vec E$ yields the same result
(41) so that recovering (32) from (33) is trivial.\par
Now the three-vectors in (33) can be decomposed in (3-space) longitudinal and
transverse parts, namely
\begin{eqnarray}
{\cal L}^{(3)} &=&  - \vec E_T \KG \partial_0 \vec A_T
- \vec E_L \KG \partial_0 \vec A_L + A^0 ( \KG \vec \nabla
\vec E_L - \rho )   \nonumber \\
 & & \mbox{} - \frac{1}{2} [ \vec E_T \KG \vec E_T + \vec E_L \KG \vec E_L
+ (\vec \nabla \times \vec A_T ) \KG (\vec \nabla \times \vec A_T ) ]
+ \vec A \vec j
\end{eqnarray}
Thus $A^0$ yields a constraint that can be solved giving
$$
\vec E_L = \KG^{-1} \Delta^{-1} \vec \nabla \rho
$$
Then the term
$$ - \vec E_L \KG \partial_0 \vec A_L = -
( \Delta^{-1} \vec \nabla \rho ) \partial_0 \vec A_L $$
can be absorbed
by the same redefinition of the fermion fields given in ref.[4], whilst one has
$$
\vec E_L  \KG \vec E_L = - \rho \KG^{-1} \Delta^{-1} \rho
$$
Because of the occurrence of the differential operator $\KG$ ,
it is not easy to read the remaining actual d.o.f directly out of (34).
This analysis just shows that the phase-space is further deprived of two
(non-covariant) variables. \par
To work out the remaining phase space we go back momentarily
to the covariant treatment and perform first the
Legendre transformation that leads to (19) and then to (21), though we already
know that they contain some physically irrelevant d.o.f.
The diagonalization (20) just rearranges the d.o.f without altering them,
and for a conserved source there is no need of further field redefinitions.\par
Now we consider the $ \tilde A $ and $ \tilde \pi $ sectors of (21)
(equation (23) is equally suited, so the choice is a matter of taste).
Both fields feature the well-known Lorentz-longitudinal parts that are
decoupled from the conserved matter source. Thus for the massless field
$\tilde A$
one is left with its Lorentz-transverse part, described by an ordinary
gauge-invariant Lagrangian as in (40).
In ref.[4] the constraint was solved showing that one finally has the
two d.o.f of a photon. For the remaining massive transverse poltergeist
$ \tilde \pi$ the analogous analysis is even simpler.
We discuss this case
in the Appendix where we will find three d.o.f as expected.\par
We finally stress that when the source is not conserved, as it would be the
case of a non-Abelian quantum theory, the two Lorentz-longitudinal parts
{\it do} couple to the matter, but this is compensated by the
(higher-derivative) FP sector of the theory.

\section{Gauge invariance and gauge fixings }
For $\zeta = 0 $ and conserved source the starting HD theory (1) is exactly
invariant under the U(1) gauge transformations (2).\par
For arbitrary $\zeta $ the variation (2) induces the following ones in the
intermediate field variables:

\begin{eqnarray}
\delta \hat A_{\mu} = \zeta \partial_{\mu} \lambda \qquad \qquad \qquad &
\delta \tilde A_{\mu} = \zeta (1+ \frac{\Box}{m^2} )  \partial_{\mu} \lambda
\nonumber \\
\delta \hat \pi_{\mu} = \zeta (\frac{1}{2} + \frac{\Box}{m^2} )
\partial_{\mu} \lambda \qquad \qquad & \delta \tilde \pi_{\mu} = -
\frac{\zeta}{m^2} \Box  \partial_{\mu} \lambda
\end{eqnarray} and in the final variables
\begin{eqnarray}
\delta {\cal A}_{\mu} &= &(1+ \frac{\Box}{m^2} )  \partial_{\mu} \lambda
\nonumber \\
\vspace{0.5 cm}
\delta \Pi_{\mu} &= & - \frac{\Box}{m^2} \partial_{\mu} \lambda
\end{eqnarray}
so that
\begin{equation}
\delta ({\cal A}_{\mu} + {\Pi}_{\mu}) = \partial_{\mu} \lambda
\end{equation}
For $\zeta = 0 $ the final LD theory (23) is therefore invariant under
the induced variations (36). Notice that the symmetric limit poses
no problem with the potentially troublesome source terms in (13),(17)
and (21). Then they actually are $j \hat A $ , $j \hat A $  and
$j( \tilde A + \tilde \pi )$ respectively because of source
conservation, and the redefinition (22) is not needed. However the
theory is also invariant under the independent variations
\begin{eqnarray}
\delta {\cal A}_{\mu} &= & \partial_{\mu} \Lambda \nonumber \\
\delta {\Pi}_{\mu} &= & \partial_{\mu} \Lambda'
\end{eqnarray}
This shows that in the LD theory one actually has a larger
accidental $ U(1) \times U'(1) $ symmetry, which is hidden in the
HD theory. If the matter source is not conserved, the diagonal
subgroup of transformations
\begin{eqnarray}
\delta {\cal A}_{\mu} &= &\partial_{\mu} \Lambda \nonumber \\
\delta {\Pi}_{\mu} &= & - \partial_{\mu} \Lambda
\end{eqnarray}
still survives. The original symmetry, as given by (36),
appears also as a subgroup. \par

However one may choose to put only the HD GF term in(1) off.
This is accomplished by taking the limit $M \rightarrow
\infty $ . What happens at the level of the propagators is now
clear in either (9) or (10) : the propagation of the "third
ghost", of the massive FP d.o.f, and of the B field fade away.
This is equivalent to suppressing the term
corresponding to the third ghost in (23). In that case the
symmetry $U'(1)$ is preserved.\par
We conclude that the LD and HD GF terms are associated to the
breaking of the $U'(1)$ and $U(1)$ symmetries respectively.
However we cannot adopt a pure HD GF because then the Legendre
transform becomes singular, as it does happen for a theory with
only HD terms.\par
\vskip 1 cm
These results are of quite a limited relevance since they are characteristic
to the Abelian theory. In fact the model we have considered describes a
theory which, except for a spectator interaction with external source
fields, is essentially free and hence trivial. Because of the occurrence of
self-interactions, the second ``hidden" symmetry is absent as soon as
a non-Abelian generalization of the model is considered. However the results
above still hold for the trivial Abelian symmetry of the free parts.

\section{ Conclusions}
{}From a HD theory of one vector field $A_{\mu} $ with quartic
propagator we have obtained an equivalent LD theory
with quadratic propagators for positive norm and poltergeist
states. Except for the gauge fixing the original theory had a
U(1) gauge symmetry.\par
Exploring the symmetries of the LD theory we have deduced that
the possible LD and HD gauge fixing terms in the HD theory
actually fix separate $U(1)$ and $U'(1)$ hidden symmetries of the free
theory. The family of compensating Faddeev-Popov ghosts
includes, also in the non-Abelian case, a further massive FP d.o.f
and a propagating massive
commuting field. The loss of unitarity due to the negative-norm
poltergeists is instead unavoidable at this level.\par
The above procedures extrapolate naturally to the theory of
gravitation with little modification up to technical details.
Beside the graviton there we have a massive spin-two poltergeist,
and a physical
scalar field in the HD Diff-invariant theory together with a richer
set of orthogonal projectors. They include a spin one and a further
scalar components when the gauge is fixed. The findings above regarding
the symmetries of the free parts,
may help to understand how a Fierz-Pauli kinetic term arises for the massive
spin-two poltergeist in the LD equivalent theory.
Work is in progress in that direction.\par
The example we have worked out in this paper is interesting also
by itself: it contains a sort of regularization that
allows an elegant proof \cite{5} of the Adler-Bardeen theorem.
 In fact this regularization, extended to the fermion propagators,
works for the potentially anomalous graphs of two or more loops \cite{6}
and preserves the chiral gauge invariance.\par

\appendix
\section*{Appendix}

We use the notations
\begin{eqnarray*}
j^{\mu} & = (\rho , \vec j ) \\
A^{\mu} & = (A^0 , \vec A ) \\
\partial_{\mu} & = ( \partial_0 , \vec \nabla )\\
\Delta & \equiv \nabla^2 \\
\Box & = \partial_0^2 - \Delta
\end{eqnarray*}
The (second order) electromagnetic Lagrangian
\begin{eqnarray}
{\cal L}^{(2)}   & = & - \frac{1}{4} \Fg  \Fs - j_{\mu} A^{\mu}  \nonumber \\
& = & \frac{1}{2} [( - \partial_0 \vec A - \vec \nabla A^0 )^2 -
(\vec \nabla \times \vec A )^2 ] - j_{\mu} A^{\mu}
\end{eqnarray}
can be cast in a first order equivalent form by a Legendre transformation.
 In fact one may define the conjugate variable
\begin{equation}
- \vec E = \frac{\partial {\cal L}^{(2)}}{\partial \partial_0 \vec A } =
\partial_0 \vec A + \vec \nabla A^0
\end{equation}
The Hamiltonian is then
\begin{equation}
{\cal H}= \frac{1}{2} \vec E^2 + \vec E \vec \nabla A^0 +
(\vec \nabla \times \vec A )^2  + j_{\mu} A^{\mu}
\end{equation}
so that the first order (Helmholtz) Lagrangian becomes
\begin{equation}
{\cal L}^{(1)} =  \vec E ( - \partial_0 \vec A - \vec \nabla A^0 ) -
\frac{1}{2} [ \vec E^2 + (\vec \nabla \times \vec A )^2 ] - j_{\mu} A^{\mu}
\end{equation}
which is equation (15a) in ref.[4], with $\vec B = \vec \nabla \times \vec A $.
Equation (40) can be recovered by solving the equation of motion for $\vec E$
in (43), obtaining (41), and then substituting it back in (43). The field
$A^0$ gives rise to a constraint that can be solved as shown in ref.[4].
This eliminates $\vec A_L $ and $\vec E_L$
from the radiating field so that only the two d.o.f of the photon survive.\par
The massive case can be treated along the same lines. The second order Proca
Lagrangian
\begin{equation}
{\cal L}^{(2)}_m   = - \frac{1}{4} \Fg  \Fs + \frac{m^2}{2} A_{\mu} A^{\mu}
- j_{\mu} A^{\mu}
\end{equation}
only adds non-derivative terms to the massless case above and contains also
a decoupled Lorentz-longitudinal d.o.f. Now the first order
Lagrangian is
\begin{equation}
{\cal L}^{(1)}_m =  \vec E ( - \partial_0 \vec A - \vec \nabla A^0 ) -
\frac{1}{2} [ \vec E^2 + (\vec \nabla \times \vec A )^2  - m^2 ( A_0^2 -
\vec A^2 )] - A^0 \rho + \vec A \vec j
\end{equation}
Again the time derivative of $A^0$ is absent but this does not yield a
constraint.
In fact, dropping total derivatives, the terms containing $A^0$ are
\begin{equation}
A^0 ( \vec \nabla \vec E_L - \rho ) + \frac{m^2}{2} {A^0}^2
\end{equation}
Then $A^0$ can be solved in terms of $\vec E_L$ and $\rho$. Substituting
it back in (45) one obtains
\begin{equation}
{\cal L}^{(1)}_m = - \vec E  \partial_0 \vec A -
\frac{1}{2m^2} ( \vec \nabla \vec E_L - \rho )^2 - \frac{1}{2}
[ \vec E^2 + (\vec \nabla \times \vec A )^2  - m^2 \vec A^2 ]
+ \vec A \vec j
\end{equation}
One sees that now $\vec E_L$ is an independent field and $\vec A_L$ cannot
be absorbed into a redefinition of the fermion fields in the matter source.
 Thus the space-longitudinal d.o.f survives and consequently we are left
with 3 d.o.f.

\section*{Acknowledgements}
We thank Prof. M.Tonin and Prof. R.Balbinot for useful remarks.\par
One of us (J.J.) wishes to thank, for the hospitality extended to him,
S.Bergia and the Theoretical Group of the Dipartimento di Fisica
dell'Universit\`a di Bologna, where this work
was done.\par

\end{document}